\documentclass[useAMS,usenatbib]{mn2e}

\def\gsim{ \lower .75ex \hbox{$\sim$} \llap{\raise .27ex \hbox{$>$}} }
\def\lsim{ \lower .75ex \hbox{$\sim$} \llap{\raise .27ex \hbox{$<$}} }
\def\Mo{{\rm M_\odot}}

\title[Cluster density profiles]
{Convergence and scatter of cluster density profiles}

\author[J\"urg Diemand, Ben Moore $\&$ Joachim Stadel]
{J\"urg Diemand\thanks{diemand@physik.unizh.ch},
Ben Moore $\&$  Joachim Stadel\\
$$Institute for Theoretical Physics, University of Z\"urich,
Winterthurerstrasse 190 ,CH-8057 Z\"urich, Switzerland}

\begin{document}

\pagerange{\pageref{firstpage}--\pageref{lastpage}} \pubyear{2004}

\maketitle

\label{firstpage}  

\begin{abstract}
We present new results from a series of $\Lambda$CDM simulations of cluster mass 
halos resolved with high force and mass resolution. These results are compared 
with recently published simulations from groups using various 
codes including PKDGRAV, ART, TPM, GRAPE and GADGET. 
Careful resolution tests show that with 25 million particles within the 
high resolution region we can resolve to about 0.3\% of the virial radius and that
convergence in radius is proportional to the mean interparticle separation.
The density profiles of 26 high resolution clusters obtained with the different 
codes and from different initial conditions agree very well. The average 
logarithmic slope at one percent of the virial radius 
is $\gamma = 1.26$ with a scatter of $\pm 0.17$.
Over the entire resolved regions the density profiles are well fitted by 
a smooth function that asymptotes to a central cusp $\rho \propto r^{-\gamma}$, where 
we find $\gamma=1.16\pm 0.14$ from the mean of the fits to our six highest 
resolution clusters. 
\end{abstract}

\begin{keywords}
methods: N-body simulations -- methods: numerical --
dark matter --- galaxies: haloes --- galaxies: clusters: general
\end{keywords}

\section{Introduction}

A highly motivated and well defined problem in computational astrophysics is to 
compute the non-linear structure of dark matter halos. This is especially timely 
given 
the abundance of new high resolution data that probe the central 
structure of galaxies (e.g. \citealt*{deBlok2001a}; 
\citealt*{deBlok2001b}; \citealt*{McGaugh2001}; 
\citealt*{Swaters2003}; \citealt*{deBlok2002}; \citealt*{Gentile2004}) 
and clusters (e.g. \citealt*{Sand2003}). Furthermore, a standard 
cosmological paradigm has been defined that gives a well defined framework
within which to perform numerical calculations of structure formation 
(e.g. \citealt{Spergel2003}). 
This subject has developed rapidly over the past few years, building 
upon the pioneering results obtained in the early 1990's by \citet{Dubinski1991}
and \citet{Warren1992}. 
More recently, the systematic study of many halos at a low resolution led to
the proposal that the profile of an `average' cold dark matter halo
in dynamical equilibrium could be fit by an universal
two parameter function \citep*{Navarro1996}, with a slope of that
asymptotically approaches $-1$ as $r \to 0$.
At the same time, the study of a few 
halos at high resolution questioned these results (\citealt*{Fukushige1997};
\citealt*{Moore1998}; \citealt*{Moore1999}; \citealt*{Jing2000}; 
\citealt*{Ghigna2000}). 
These latter authors claimed that of the order
a million particles within the virialised region where necessary to resolve the 
halo structure to 1\% and the slopes at that radius could be significantly 
steeper.
Just within the last few months, we have seen several 
groups publish reasonably large samples of halos simulated with the 
necessary resolution that we can finally determine the 
scatter in the  density profiles across a range of mass scales 
(\citealt*{Fukushige2003}; \citealt*{Tasitsiomi2003}; \citealt*{Wambsganss2003};
 \citealt*{Hayashi2003}; \citealt*{Navarro2003}; \citealt*{Reed2003}). 

Much of the recent controversy in the literature has been due to limited 
statistics and the lack of agreement over what is a reliable radius to trust
a given simulation with a given set of parameters. 
Several studies have attempted to address this issue 
(\citealt*{Moore1998}; \citealt*{Knebe2000}; \citealt*{Klypin2001};
\citealt*{Power2003}; \citealt*{Diemand2004}).
Integration and force accuracy can be understood using controlled test 
simulations.
However, discreteness is probably the most important and least understood 
numerical 
effect that can influence our numerical results which is exacerbated due to the 
lack of 
an analytic solution with which to compare simulations.  
Our particle sampling of the nearly collisionless fluid
we attempt to simulate can lead to energy transfer and mass redistribution, 
particularly in the central regions that we are often most interested in.

Collisional effects in the final object or in the early hierarchy of objects can 
be reduced by 
increasing the number of particles $N$ in a simulation \citep{Diemand2004}. 
The limitation to the phase space 
densities that can be resolved due to discreteness in the initial conditions can 
also be overcome by 
increasing the resolution \citep{Binney2003}. As we increase the resolution within
a particular non-linear structure, we 
find that the global 
properties of the resolved structure is retained, including shape, density 
profile, 
substructure mass functions and even the positions of the infalling 
substructures. 
This gives us confidence that our $N$-body calculations are not biased by 
using finite $N$ \citep{Baertschiger2002}. 
The fact that increasing the resolution allows us to resolve smaller 
radii is important 
since the 
baryons often probe just the central few percent of a dark matter structure - 
the 
latest observations of galaxy and clusters probe the mass distribution within 
one percent
of the virial radius, which until
recently was unresolved by numerical simulations. Forthcoming experiments, such 
as 
VERITAS \citep{Weekes2002} and MAGIC \citep*{Flix2004} will 
probe the structure of dark matter halos on even smaller scales by attempting 
to detect gamma-rays from dark matter annihilation within the central hundred 
parsecs ($\sim 0.1\%R_{virial}$) of the Galactic halo (e.g. \citealt*{Calcaneo2000}). 
These scales are still below the resolution limit of todays cosmological simulations, 
the estimates of the dark matter densities in these regions are still based on extrapolations
which introduce large uncertainties

A simple estimate of the scaling of $N$ with time shows remarkable progress over 
and above that predicted by Moore's law. The first computer simulations used 
of the order $10^2$ particles and force resolution of the order of the half mass 
radii \citep{Peebles1970}. 
Today we can follow up to $10^8$ particles with a resolution of $10^{-3}$
of the final structure. The increase in resolution is significantly faster than 
predicted by
Moore's law since equally impressive gains in performance have been due to 
advances in 
software.

We are finally at the stage whereaby the dark matter clustering is understood at 
a level where the uncertainties are dominated by the influence of the baryonic 
component. It is therefore a good time to review and compare existing results 
from different groups 
together with a set of new simulations that we have carried out that are the 
state of the art
in this subject and represent what is achievable with several months of 
dedicated 
supercomputer time. For certain problems, such as predicting the annihilation 
flux 
discussed earlier, it would be necessary to significantly increase the 
resolution. This is not possible with existing resources and new 
techniques should be explored. We begin by presenting our new simulations 
in Section 2. Section 3 discusses convergence tests and the asymptotic best
fit density profiles. In Section 4 we compare our results with 
recently published results from four other groups mentioned above.

\section{Numerical Experiments}\label{Sim}

Table \ref{tab1} gives an overview of the simulations we present in this paper.
With up to $25\times 10^6$ particles inside the virial radius of a cluster
and an effective $10^5$ timesteps, they are among the highest resolution 
$\Lambda$CDM
simulations performed so far. They represent a major investment of computing 
time, the largest run was completed in about $10^5$ CPU hours on the zBox 
supercomputer \footnote{http://www-theorie.physik.unizh.ch/$\sim$stadel/zBox/}.

\begin{table*}
\centering
\begin{minipage}{140mm}
\caption{Parameters of simulated cluster halos}
\label{tab1}
\begin{tabular}{l|c|c|c|c|c|c|c|c|c}
\hline
Run&$z_i$&$\epsilon_0$&$\epsilon_{\rm max}$&$N_{\rm vir}$&
$M_{\rm vir}$&$r_{\rm vir}$&$Vc_{\rm max}$&$r_{\rm max}$&$r_{\rm resolved}$\\
 & &[kpc]&[kpc]& &$10^{15} [\Mo]$&[kpc]&[km s$^1$]&[kpc]&[kpc]\\
 \hline
 $A9$& 40.27& 2.4 & 24 & 24'987'606 & 1.29 & 2850 & 1428 & 1853 & 9.0 \\
 $B9$ & 40.27 & 4.8 & 48 & 11'400'727 & 0.59 & 2166 & 1120 & 1321 & 14.4 \\
 $C9$ & 40.27 & 2.4 & 2.4 & 9'729'082 & 0.50 & 2055 & 1090 & 904 & 9.0 \\ 
 &&&&&&&&&\\
 $D3h$ & 29.44 & 1.8 & 18 & 205'061& 0.28 & 1704 & 944 & 834 & 27 \\
 $D6h$ & 36.13 & 1.8 & 18 & 1'756'313 & 0.31 & 1743 & 975 & 784 & 13.5 \\
 $D6$ &  36.13 & 3.6 & 36 & 1'776'849 & 0.31 & 1749 & 981 & 840 & 13.5 \\
 $D9$ & 40.27 & 2.4  & 24  & 6'046'638 & 0.31 & 1752 & 983 & 876 & 9.0 \\
 $D9lt$ & 40.27 & 2.4 & 24 & 6'036'701 & 0.31 & 1752 & 984 & 841 & 9.0 \\
 $D12$ & 43.31 &1.8 & 18 & 14'066'458 & 0.31 & 1743 & 958 & 645 & 6.8 \\
 &&&&&&&&&\\ 
 $E9$ & 40.27 & 2.4 & 24 & 5'005'907&  0.26 & 1647 & 891 & 889 & 9.0 \\
 &&&&&&&&&\\
 $F9$ & 40.27& 2.4 & 24 & 4'567'075 & 0.24 & 1598 & 897 & 655 & 9.0 \\
 $F9cm$ & 40.27 & 2.4 & 2.4 & 4'566'800 & 0.24 & 1598 & 898 & 655 & 9.0 \\
 $F9ft$ & 40.27& 2.4 & 99.06 & 4'593'407 & 0.24 & 1601 & 905 & 464 & 9.0 \\
\end{tabular}
\end{minipage}
\end{table*} 

\subsection{N-body code and numerical parameters}\label{Code}

The simulations have been performed using a new version of 
PKDGRAV, written by Joachim Stadel and Thomas Quinn \citep{Stadel2001}. 
The code was optimised to reduce the computational cost of the very high 
resolution runs we present in this paper. We tested the new version
of the code by rerunning the ``Virgo cluster'' 
initial conditions \citep{Moore1998}.
We confirmed that density profile, shape of the cluster and 
the amount of substructure it contains is identical 
to that obtained with the original code presented in \citet{Ghigna1998}.

Individual time steps are chosen for each particle proportional to the square 
root of the softening length over the acceleration, 
$\Delta t_i = \eta\sqrt{\epsilon/a_i}$. We use $\eta = 0.2$ for most runs,
only in run $D9lt$ we used larger timesteps $\eta = 0.3$ for comparison. 
The node-opening angle is set to $\theta = 0.55$ initially, and after $z = 2$
to $\theta = 0.7$. 
This allows higher force accuracy when the mass distribution
is nearly smooth and  the relative force errors can be large in the treecode.
Cell moments are expanded to fourth order in PKDGRAV, other treecodes
typically use just second or first order expansion.
The code uses a spline 
softening length $\epsilon$, forces are completely Newtonian at $2\epsilon$.
In Table \ref{tab1} $\epsilon_0$ is the softening length at  $z=0$, 
$\epsilon_{max}$ is the maximal softening in comoving coordinates.
In most runs the softening is constant in physical coordinates from $z=9$ to the 
present and is constant in comoving coordinates before,
i.e. $\epsilon_{\rm max} = 10 \epsilon_0$. In runs $C9$ and $F9cm$ the softening 
is constant in comoving coordinates for the entire run, in run $F9ft$ the
softening has a constant physical length for the entire run.
 
\subsection{Initial conditions and cosmological parameters}\label{IC}

We adopt a $\Lambda$CDM cosmological model with parameters from the first year 
WMAP results:
$\Omega_{\Lambda} = 0.732$, $\Omega_m = 0.268$, $\sigma_8 = 0.9$, $h = 0.71$, 
\citep{Spergel2003}.
The initial conditions are generated with the 
GRAFIC2 package \citep{Bertschinger2001}.
The starting redshifts $z_i$ are set to the time when 
the standard deviation of the density fluctuations 
in the refined region reaches $0.2$.

First we run a parent simulation: a $300^3$ particle cubic grid with
a comoving cube size of $300$ Mpc (particle mass $m_{\rm p} = 3.7\times 10^{10} 
\Mo$, force resolution $\epsilon_0 = 100 kpc$, $\epsilon_{\rm max}  = 1 Mpc$).
Then we use the friends-of-friends (FoF)
algorithm \citep{Davis1985} with a linking length of $0.164$ mean interparticle 
separations to
identify clusters. We found 39 objects with virial masses above $2.3\times 
10^{14} \Mo$.
We selected six of these clusters 
for resimulation,
discarding objects close to the periodic boundaries and objects that show clear 
signs of recent major mergers at $z=0$.
We label the six cluster with letters $A$ to $F$ according to their mass.
It turned out that two of the clusters selected in this way (runs $A$ and $C$) 
have ongoing major mergers at $z=0$ (i.e. two clearly distinguishable central cores),
which is not evident from the parent simulation due to lack of resolution. These 
clusters were evolved slightly into the future to obtain 
a sample of six 'relaxed' clusters.

For re-simulation we mark and trace back the particles within a cluster's virial radius
to the initial conditions. All particles which lie within a $4$ Mpc (comoving) thick
region surrounding the marked particles in the initial conditions
are also added to the refinement region. 
This ensures that there is no pollution of heavier particles
within the virial radius of the resimulated cluster. Typically one third or one 
quarter of the refinement particles ends up within the virial radius.
To reduce the mass differences at the border of
the refinement region we define a $5$ Mpc thick 'buffer region' around the high 
resolution region, there an intermediate refinement factor of 3 or 4 in length 
is used.
The final refinement factors are $6$, $9$ and $12$ in length, 
i.e., $216$, $729$ and $1728$ in mass, 
so that the mass resolution is $m_{\rm p} = 2.14 \times 10^7 \Mo$ in the highest 
resolution run. We label each run with a letter indicating the object and number 
that gives the refinement factor in length. To reduce the mass differences at 
the 
border of
the refinement region we define a $5$ Mpc thick 'buffer region' around the high 
resolution region, there an intermediate refinement factor of 3 or 4 in length 
is used.
 
\subsection{Measuring density profiles}

We define the virial radius $r_{\rm vir}$ such that the mean density within 
$r_{\rm vir}$ is $178 \Omega_M^{0.45} \rho_{\rm crit} = 
98.4 \rho_{\rm crit}$ for the adopted model \citep*{Eke1996}.
We use 30 spherical bins of equal logarithmic width, centered on the densest
region of each cluster using TIPSY
\footnote{TIPSY is available form the University
of Washington N-body group: 
http://www-hpcc.astro.washington.edu/tools/tipsy/tipsy.html}.
We confirmed that using triaxial bins adapted to the shape of the
isodensity surfaces (at some given radius, we tried 
$0.1$, $0.5$ and  $1 r_{\rm vir}$) does not change the form of
the density profile, in agreement with \citet{Jing2002}.
For simplicity and easier comparison to other
results we will present only profiles obtained using spherical bins.
Data points are plotted at the arithmetic mean of the 
corresponding bin boundaries;
the first bin ends at $1.5$ kpc, the last bin at the virial radius.

\section{$\Lambda$CDM Cluster Profiles}\label{profiles}

\subsection{Profile Convergence Tests}\label{convergence tests}

Numerical convergence tests show that with sufficient timesteps, force accuracy 
and 
force resolution the
radius a CDM simulation can resolve is limited by the mass resolution
(\citealt{Moore1998}; \citealt{Ghigna2000}; \citealt{Knebe2000};
\citealt{Klypin2001}; \citealt{Power2003}; \citealt{Hayashi2003}; 
\citealt{Fukushige2003}; \citealt{Reed2003}). These tests
compare different mass resolution simulations of the same object to determine 
the resolved radius. 
The resulting radii scale with $N^{-0.45}$ according to \citet{Power2003},  
\citet{Hayashi2003} and \citet{Fukushige2003}, 
but only with $N^{-1/3}$ in the tests in \citet{Moore1998},\citet{Ghigna2000} 
and \citet{Reed2003}.

\begin{figure*}
\vskip 6.0 truein
\includegraphics{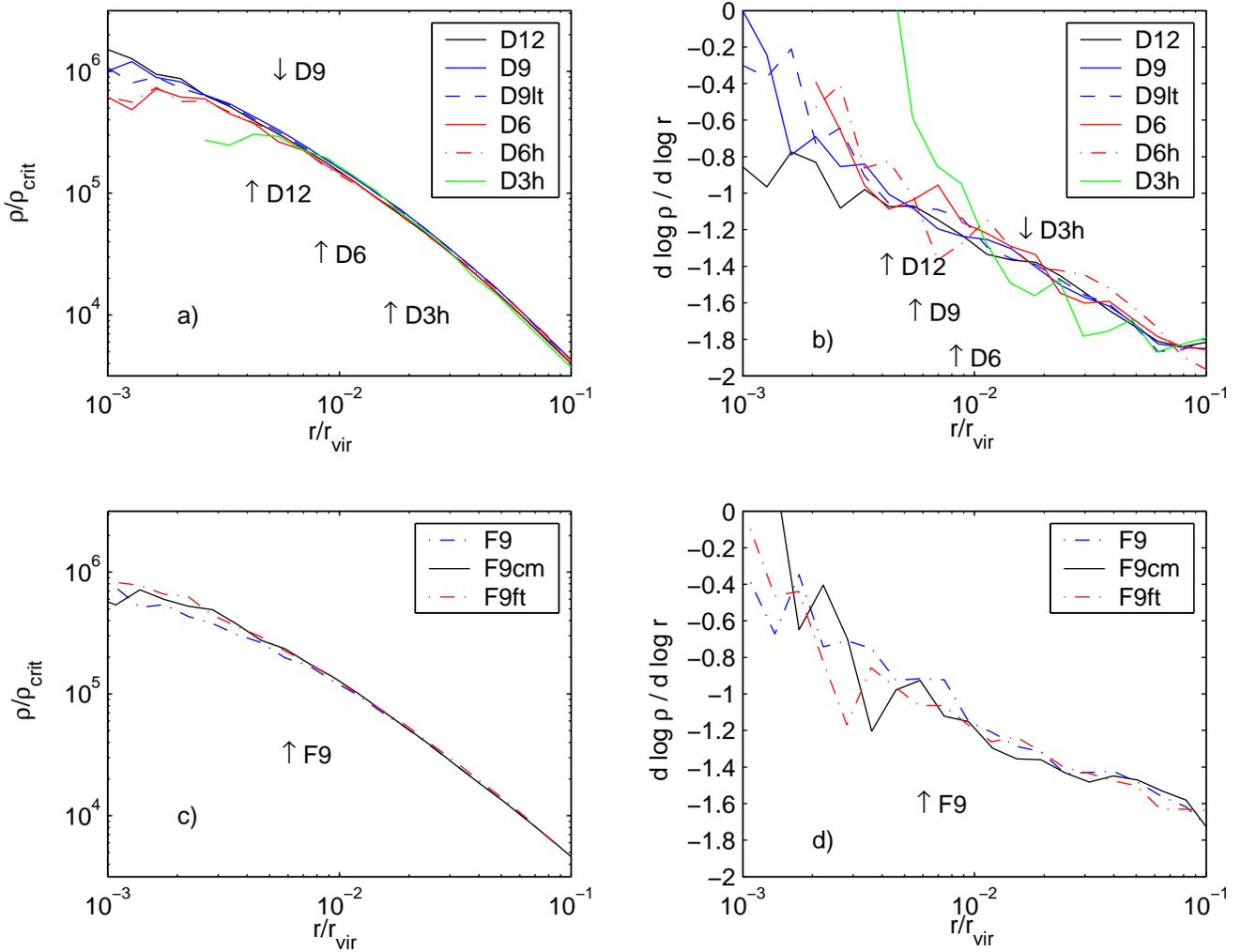}
\caption{\label{proConv.eps}Numerical convergence tests for the cluster 
profiles:
Panel (a): Density profiles of cluster $D$
resolved with $N_{\rm vir} = 205k, 1.8M, 6M$ and $14M$ particles.
Panel (b): Logarithmic slope for the profiles from (a).
Panel (c): Density profiles of cluster $F$ simulated with different numerical 
parameters: $F9ft$ used 4096 fixed timesteps and constant $\epsilon$ in physical
coordinates as in \citet{Fukushige2003}. 
$F9cm$ and $F9$ used adaptive timesteps $0.2 \sqrt{\epsilon(z)/a}$
with comoving softening in $F9$ and mixed comoving/physical softening in $F9$
($\epsilon_{\rm max} = 10 \epsilon_0$).
Panel (d): Logarithmic slope for the profiles from (c).} 
\end{figure*}

\begin{figure}
\vskip 3.2 truein
\includegraphics{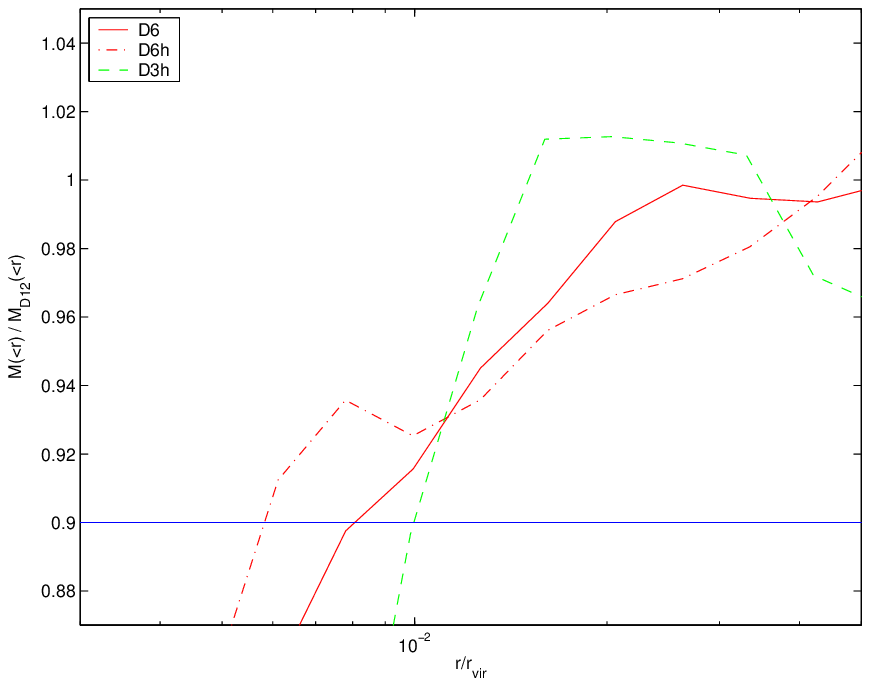}
\caption{\label{proConv2.eps} Ratios of the mass enclosed in low resolution 
runs to mass enclosed in the high resolution run $D12$. By comparing runs
with equal softening (smaller than one third of the convergence scale) like
$D3h$ and $D6h$ one finds that the resolved radii scale like $r \propto N^{-1/3}$.
A larger softening (see run $D6$) can increase the converged scales and change this 
scaling.} 
\end{figure}

\subsubsection{Mass resolution}

The finite mass resolution of N body simulations always leads to two body 
relaxation effects, i.e. heat
is transported into the cold halo cores and they expand. It is not obvious that
better mass resolution reduces the effects of two body relaxation, 
since in hierarchical models the first resolved objects always contain 
just a few particles and with higher resolution these first objects form
earlier, i.e. they are denser and more affected by relaxation effects 
(\citealt*{Moore2001}; \citealt*{Binney2002}). Estimates
of relaxation based on following the 
local phase-space density in simulations 
show that the amount of relaxation can be reduced with
better mass resolution, but the average degree of relaxation scales roughly like 
$N^{-0.3}$
much slower than $N^{-1}$ expected from the relaxation time of the final 
structure \citep{Diemand2004}.
This confirms the validity of performing convergence tests in $N$,
but one has to bear in mind that convergence can be quite slow.

\begin{table}
\caption{Convergence radii measured by comparing with run $D12$.
The numbers in the run labels are $\propto N^{1/3}$, at fixed force
resolution we get $r \propto N^{-1/3}$ (bold values). 
Question marks indicate that a run with much better
mass resolution than $D12$ would be needed to measure this convergence radii 
reliably.
Stars indicate estimated radii assuming a convergence rate 
of $r \propto N^{-1/3}$.}
\label{tabConv}
\begin{tabular}{l | c | c | c | c | c | c | c | c }
  \hline
  Run &$\epsilon_0$&$N_{\rm vir}$&$r_{\rm 10\%vc}$&$r_{\rm 10\%M}$&$r_{\rm 
10\%\rho}$\\
&[kpc]& &[kpc]& [kpc]&[kpc]\\
  \hline
  $D3h$ & 1.8 & 205'061&{\bf 17.2}&{\bf 21.9}&{\bf 9.5}\\
  $D6h$ & 1.8 & 1'756'313&{\bf 8.4}&{\bf 10.7}&{\bf 4.6}\\
  $D6$ & 3.6 & 1'776'849 & 8.4 & 17.3 & 12.1\\
  \hline
  $D9$ & 2.4  & 6'046'638 & 3.2 ? & 5.2 ? & 2.2 ?\\
  $D9lt$ & 2.4 & 6'036'701 & 5.2 ? & 6.6 ? & 2.8 ?\\
  \hline
  $D9$ & 2.4  & 6'046'638 & 5.7 * & 7.3 * & 3.2 *\\
  $D12$ &1.8 & 14'066'458 & 4.2 * & 5.3 * & 2.4 *\\
  \hline
\end{tabular}
\end{table} 

We checked a series of resimulations of the same cluster (D) for convergence in 
circular velocity, mass enclosed
\footnote{Convergence within 10\% in cumulative mass is the same as  convergence 
in circular velocity with a tolerance of 5\%} and
density. Outside of the converged radii the values must be within 10\% of the 
reference run
D12. Table \ref{tabConv} shows the measured converged radii. 

\begin{enumerate}

\item Convergence is slow, roughly $\propto N^{-1/3}$. Therefore a high 
resolution reference run should have at least 8 times
as many particles. Between run D9 and D12 the factor is only 2.37. 
Using D12 to determine the converged 
radii of D9 gives radii that are about a factor two too small (Table 
\ref{tabConv}). 
\citet{Fukushige2003} compare
runs with $N_{\rm vir} =14 \times 10^6$ and $N_{\rm vir} = 29 \times 10^6$.
At radii where both runs have 
similar densities it is still not clear if the simulations have converged, even 
higher 
resolution studies are needed to demonstrate this.

\item If one sets the force resolution to one half of expected resolved radius, 
then it is not surprising
to measure a resolved radius close to the expected value. With this method one 
can demonstrate almost arbitrary
convergence criteria, as long as they overestimate $r_{conv}$. Therefore 
convergence tests in $N$ should
be performed with small softenings (high force resolution). 
Runs D3h, D6h and D12 all have $\epsilon_0 = 
1.8$ kpc, their converged radii
scale like the mean interparticle separation $N^{-1/3}$. In run D6 $\epsilon_0 = 
3.2$ kpc is close
to the 'optimal value' from \citet{Power2003}, and the converged radii are 
larger than in D6h (see Figure \ref{proConv2.eps}).

\item Different small scale noise in the initial conditions leads to different 
formation histories. Therefore
the shape and the density profile can differ even at radii were all runs have 
converged. For example 
between r=10 kpc and 320 kpc the densities in run D9 are about 7\%  higher than 
in run D12. Therefore
the densities in D9 are within 10\% from those of D12 quite early. If one 
rescales
$\rho$ in this range $r_{\rm 10\%\rho}$ of D9 grows from 2.2 kpc to 4.6 kpc.

\end{enumerate}

Extrapolating $r_{\rm conv} \propto N^{-1/3}$ to our highest resolution runs
gives the values on the last two lines of Table \ref{tabConv}. Note that this is 
just an extrapolation, it is not clear that 
this scaling is valid down to this level, only larger 
simulations could verify this.
To be conservative we assume the limit due to mass resolution to be $9$ kpc for 
the '9-series' of runs, and $6.8$ kpc for run D12. The force resolution sets 
another
limit at about $3 \epsilon_0$ (\citealt{Moore1998},\citealt{Ghigna2000}). 
We give the larger of the two limits as the trusted radius in Table \ref{tab1}.

\subsubsection{Force and time resolution}

Finite timesteps and force resolution also sets a limiting radius/density
that a run can resolve. We use 
multistepping, individual timesteps for the particles that are obtained by 
dividing 
the main timestep (usually $t_0/200$) by two until it is smaller
than $\eta\sqrt{\epsilon(z)/a}$, where $a$ is the local acceleration. Our 
standard choice is
 $\eta=0.2$ and $\epsilon(z=0)$ between $0.001 r_{\rm vir}$ and $0.0022 r_{\rm 
vir}$, which is chosen to be less than one third of the resolution limit expected
from the finite mass resolution. $\epsilon$ is constant
in physical length units since $z=9$ and comoving before that epoch. Here we 
argue that
the resolution limit imposed by this choice of multistepping lie well below the 
scale affected 
by finite mass resolution.

In run $D9lt$ the number of timesteps was reduced by using $\eta=0.3$, at equal 
force resolution as in $D9$.
Run $F9cm$ had a constant comoving softening during the entire simulation, in 
run $F9ft$ the softening
is physical and the timesteps are fixed $\Delta t = t_0 / 4096$ and equal for 
all particles (i.e. the same
numerical parameters as in \citet{Fukushige2003}). 
The density profiles are very similar (Figure \ref{proConv.eps}, Panel c),
there is no significant difference above the mass resolution scale of $9$ kpc. 
There is a small difference
in the inner profile of $F9$ compared to $F9ft$ and $F9cm$, at large $z$ this 
run has larger $\epsilon$ and
therefore larger timesteps than $F9cm$. So it is possible that runs with our 
standard parameters have 
slightly shallower density profiles at the resolution limit than runs with 
entirely comoving softening, or runs
with a sufficiently large number of fixed timesteps. However run $F9cm$ takes 
twice as much CPU time
as run $F9$ and run $F9ft$ three times more, therefore we accept 
this compromise.

\begin{figure}
\vskip 3.2 truein
\includegraphics{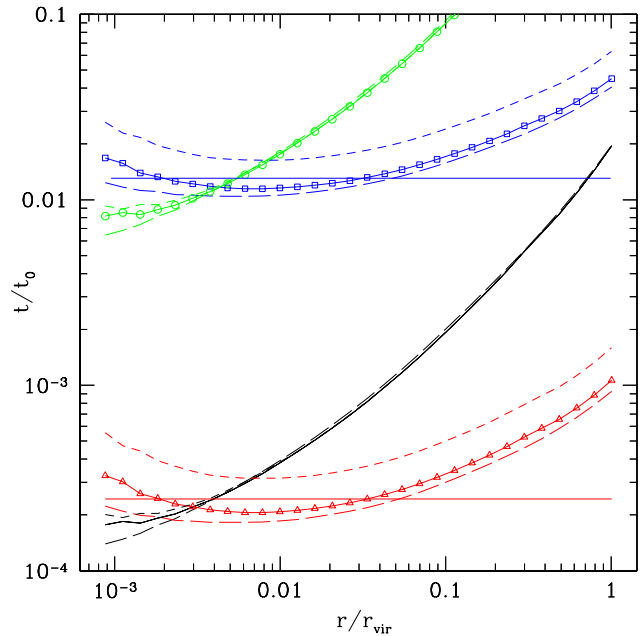}
\caption{\label{timesteps.eps}The triangles show the timestep criterium  
$\eta\sqrt{\epsilon(z)/a}$ as
a function of radius for run $D9$ at $z=0$. The dashed line is for run $D9lt$, 
which has $\eta = 0.3$, 
and the long dashed line for run $D12$. The open squares give $15 (\Delta 
t/t_0)^{5/6} t_{\rm circ} (r_{vir})$
form \citet{Power2003}, the circles are the circular orbit timescale $2\pi 
r/v_{\rm circ}(r)$. Lines
without symbols show $t_{\rm dyn}/15 = 1/(\sqrt{G\rho(<r)} 15)$. The two 
horizontal lines are 
the timesteps and $15 (\Delta t/t_0)^{5/6} t_{\rm circ} (r_{vir})$ for run 
$F9ft$.}
\end{figure}

Figure \ref{timesteps.eps} shows the timestep criterion  
$\eta\sqrt{\epsilon(z)/a}$ as
a function of radius at $z=0$ for runs $D9$ (triangles, solid line), $D9lt$ 
(dashed) $D12$ (long dashed) and for $F9ft$ (horizontal line). Particles near 
the cluster 
centre must 
take timesteps below $2\times 10^{-4}t_0$,
i.e. their timesteps are $t_0/200\times 2^{-5} = t_0/6400$. According to 
\citet{Power2003} the resolution
limit due to finite timesteps $t_{\rm ts}$ is where the circular velocity 
(circles) equals
 $15 (\Delta t/t_0)^{5/6} t_{\rm circ} (r_{vir})$ (open squares). This radius is 
indeed close to that where the circular velocities and densities start to 
differ, however for run
$D9lt$ this estimate is even a bit too conservative, since the density (and also 
$v_{\rm circ}$) 
profiles of $D9lt$ and $D9$ agree down to at least $0.005 r_{\rm vir}$. This 
suggest that about
$15$ timesteps per local dynamical time are sufficient for 
the simulations presented here. Note that this is probably not
a general condition for all cosmological simulations: Other codes seem to
require different convergence conditions than those we present in 
this paper. For example,
\citet{Fukushige2003} found that their runs converge down to $0.003 r_{\rm vir}$ 
even with only $2048$ fixed timesteps, which corresponds to only 
eight timesteps per dynamical time at this radius. 

\subsection{Density Profiles}

In this section we present the profiles of the six high resolution runs:
$A9$,$B9$,$C9$,$D12$,$E9$,$F9cm$. The output at $z=0$ was used, except for
clusters $A9$ and $C9$ which had a recent major merger
\footnote{An mpeg movie of the formation of 
cluster $C9$ can be downloaded from 
http://www-theorie.physik.unizh.ch/$\sim$diemand/clusters/}
and the core of the
infalling cluster is at about $0.02 r_{\rm vir}$ in $A9$
and at $0.1 r_{\rm vir}$ in $C9$. These cores spiral in due 
to dynamical friction and in the 'near' future both clusters have a
regular, 'relaxed' central region again. Therefore we use 
outputs at $z=-0.137$ ($+2.1$ Gyr) for run $A9$ and
$z=-0.167$ ($+2.6$ Gyr) for $C9$.

\subsection{Two parameter fits}

\begin{figure*}
\vskip 5.5 truein
\includegraphics{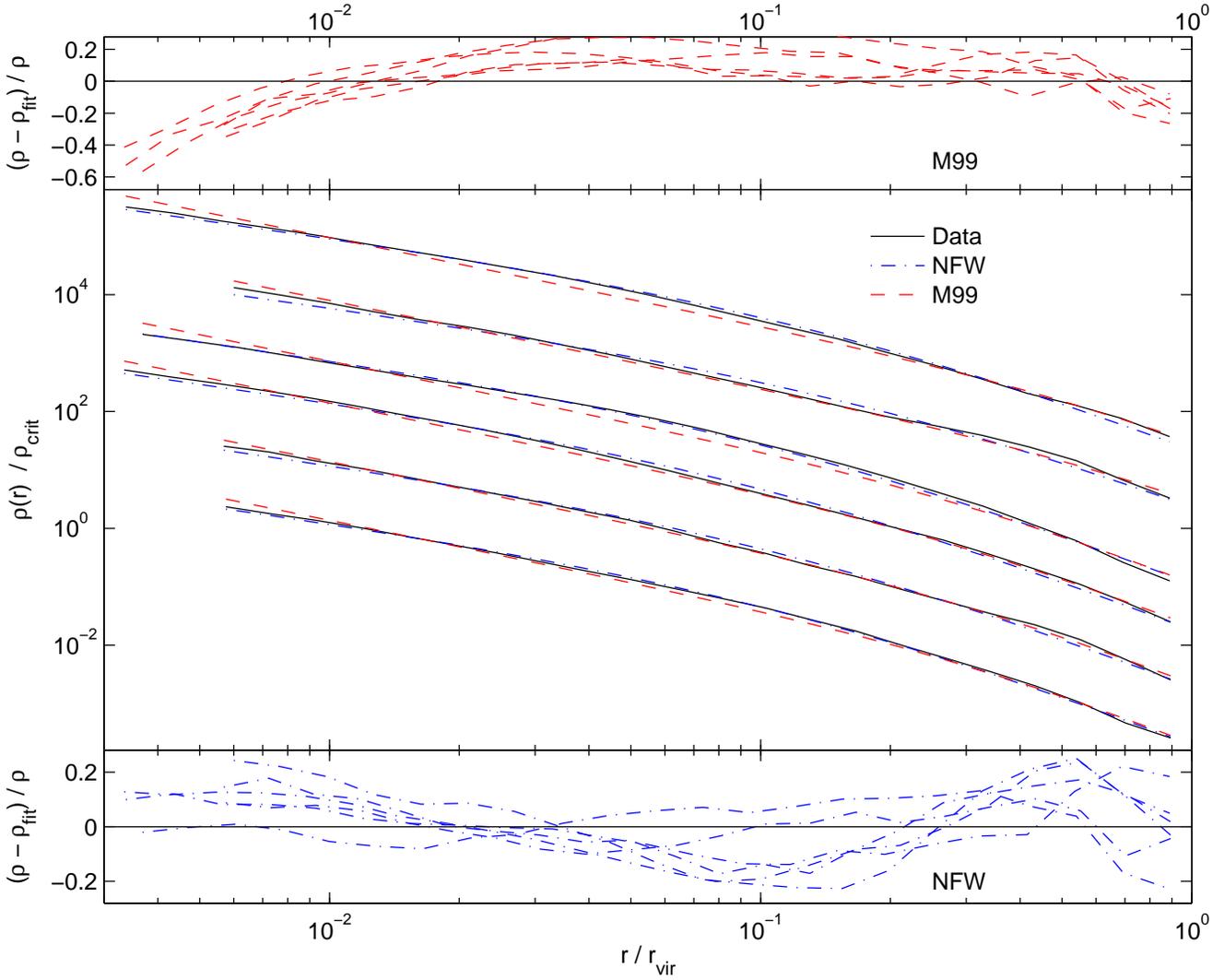}
\caption{\label{proHRfits.eps} 
Density profiles of the six clusters in our sample, clusters $B$ to $F$ are 
shifted downwards for clarity. Clusters are ordered by mass form top to bottom.
Profiles of cluster $A$ and $C$ are shown at redshifts $-0.14$ and $-0.17$, i.e.
when they have reached a 'relaxed' state with one well defined centre.
Best fit NFW and M99 profiles and residual are shown, 
obtained by minimising the squares of the relative density 
differences.} 
\end{figure*}

Figure \ref{proHRfits.eps} shows the density profiles of 
the six different clusters. We also show
best fits to functions previously proposed in the literature that have 
asymptotic central
slopes of -1 (\citealt*{Navarro1996};NFW) and -1.5 (\citealt*{Moore1999};M99). 
The fits are carried out over the resolved region by 
minimising the mean square of the relative density
differences. These two profiles have two free
parameters, namely the scale radius $r_s$ and the 
density at this radius $\rho_s = \rho(r_s)$.
The scale radii $r_s$ of these best fits  
give the concentrations $c = r_{\rm vir}/r_s$ listed in Table \ref{fitParams}.
The residuals are plotted in the top and bottom panels of Figure 
\ref{proHRfits.eps}
and the rms of the residual are given in Table \ref{fitParams} as 
$\Delta_{\rm NFW}$ and $\Delta_{\rm M99}$.
The residuals are quite large and show that neither profile is a good fit 
to all the simulations which lie somewhere in between these two extremes.

\begin{table*}
\centering
\begin{minipage}{140mm}
\caption{Density profile parameters. $\Delta$ is the root mean square of $(\rho-
\rho_{\rm fit})/\rho$
for the four fitting functions used.}
\label{fitParams}
\begin{tabular}{l | c | c || c | c || c | c || c | c | c || c | c | c }
  \hline
Run &$c_{\rm NFW}$&$\Delta_{\rm NFW}$&$c_{\rm M99}$&$\Delta_{\rm M99}$&
$\gamma_{\rm G}$&$c_{\rm G}$&$\Delta_{\rm G}$& $\alpha_{\rm N}$&$c_{\rm 
N}$&$\Delta_{\rm N}$  \\
  \hline
  $A9$ & 5.7 & 0.10 & 1.7 & 0.21 & 1.16 & 3.9 & 0.057 & 0.167 & 4.2 & 0.033 \\
  $B9$ & 4.2 & 0.16 & 1.5 & 0.13 & 1.29 & 2.1 &  0.083 & 0.141 &  2.6 & 0.093\\
  $C9$ & 7.6 & 0.09 & 3.0 & 0.26 & 0.92 & 8.7 & 0.081 & 0.247 & 7.2 & 0.068\\
  \\
  $D3h$ & 7.4 & 0.17 & 3.9 & 0.13 & 1.42 & 4.0 & 0.103 & 0.175 & 7.3 & 0.101\\
  $D6h$ & 7.9 & 0.11 & 3.8 & 0.13 & 1.17 & 4.6 & 0.089 & 0.206 & 7.2 & 0.081\\
  $D6$ & 7.9 & 0.12 & 3.8 & 0.16 & 1.25 & 5.4 & 0.101 & 0.193 & 7.2 & 0.097\\
  $D9$ & 8.8 & 0.12 & 3.9 & 0.12 & 1.21 & 6.2 & 0.096 & 0.190 & 7.8 & 0.087\\
  $D9lt$ & 8.7 & 0.12 & 3.8 & 0.12 & 1.20 & 6.2 & 0.098 & 0.191 & 7.7 & 0.087\\
  $D12$ & 8.4 & 0.12 & 3.1 & 0.14 & 1.25 & 4.5 & 0.066 & 0.174 & 6.9 & 0.051\\
  \\
  $E9$ & 7.4 & 0.12 & 3.0 & 0.10 & 1.25 & 4.5 & 0.072 & 0.176 & 6.2 & 0.069\\
  \\
  $F9$ & 6.9 & 0.06 & 3.0 & 0.14 & 1.02 & 6.7 & 0.054 & 0.224 & 6.5 & 0.048\\
  $F9cm$ & 7.3 & 0.06 & 3.1 & 0.14 & 1.10 & 6.2 & 0.055 & 0.212 & 6.6 & 0.057\\
  $F9ft$ & 7.2 & 0.05 & 3.1 & 0.16 & 1.05 & 6.6 & 0.043 & 0.218 & 6.5 & 0.045\\
  \hline   
\end{tabular}
\end{minipage}
\end{table*}

\subsection{Three parameter fits}

\begin{figure*}
\vskip 5.5 truein
\includegraphics{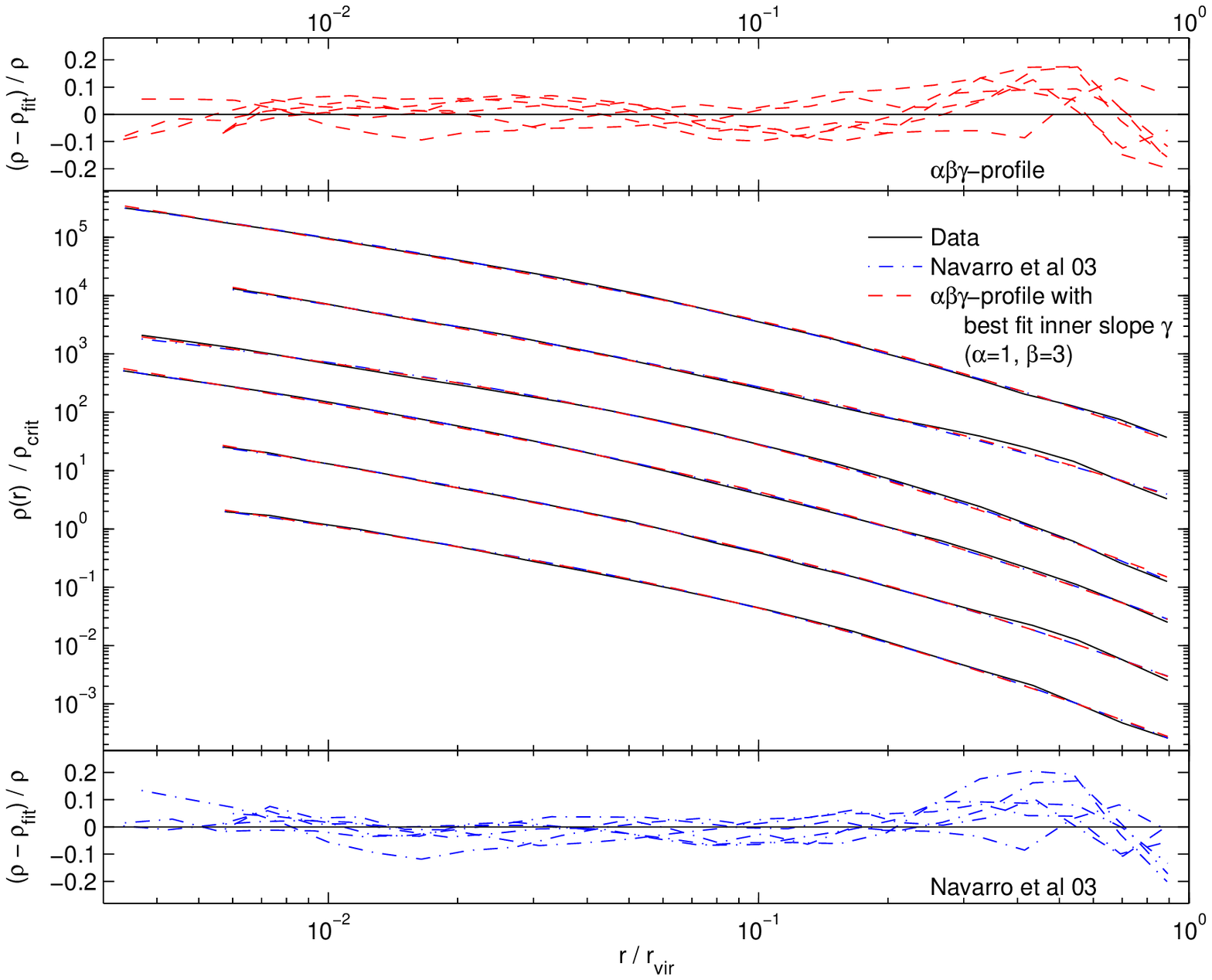}
\caption{\label{proHRfitsG1.eps} 
Same as Figure \ref{proHRfits.eps}, but with fitting functions that have
one additional free parameter. The dashed dotted lines show the profile 
(\ref{Npro}) proposed by \citep{Navarro2003}. 
The dashed lines show a general $\alpha\beta\gamma$-profile (\ref{Gpro}).
We fitted the inner slope $\gamma$ to the data and used fixed values for the 
outer slope $\beta = 3$ and turning parameter $\alpha = 1$. 
$\gamma=1$ corresponds to the NFW profile. The fit parameters and rms
of the residuals are given in Table \ref{fitParams}.}
\end{figure*}

\cite{Navarro2003} argue the large residuals of NFW and M99 fits
are evidence against any constant asymptotic 
central slope and propose a profile which curves smoothly over 
to a constant density at very small radii:
\begin{equation}\label{Npro}
\ln(\rho_{\rm N}(r)/\rho_s) = 
(-2/\alpha_{\rm N}) \left[ (r/r_s)^{\alpha_{\rm N}}  - 1 \right] \,
\end{equation}
This function gives a much better fit to the simulations, see the 
dashed dotted lines in Figure \ref{proHRfitsG1.eps},
but this should be expected since there is an additional third free parameter
$\alpha_{\rm N}$, while the NFW and M99 profiles only have two free
parameters. $\alpha_{\rm N}$ determines how fast the profile (\ref{Npro}) 
turns away from an power law near the centre. \cite{Navarro2003} found that
$\alpha_{\rm N}$ is independent of halo mass and
$\alpha_{\rm N} = 0.172 \pm 0.032$ for all their simulations, including 
galaxies and dwarfs.
The mean and scatter of our six high resolution clusters 
is $\alpha_{\rm N} = 0.186 \pm 0.037$. (Excluding cluster $C9$
yields $\alpha_{\rm N} = 0.174 \pm 0.025$). 

We also show fits to a general $\alpha\beta\gamma$-profile \citep*{Zhao1996} 
($\rho_{\rm G}$, subscript 'G' stands for 'general')
that asymptotes to a central cusp $\rho(r) \propto r^{-\gamma}$:  
\begin{equation}\label{Gpro}
\rho_{\rm G}(r) = \frac{\rho_s}{(r/r_s)^{\gamma}(1 + (r/r_s)^{\alpha})^{(\beta-
\gamma)/\alpha} } \,.
\end{equation}
We fix the outer slope $\beta = 3$ and the turnover parameter $\alpha = 1$. 
For comparison the NFW profile has $(\alpha, \beta, \gamma) = (1,3,1)$, the M99 
profile 
has $(\alpha, \beta, \gamma) = (1.5,3,1.5)$. We fit the three parameters
$\gamma$, $r_s$ and $\rho_s$ to the
data and find that this cuspy profile also provides a very good fit to the data.
The best fit values and rms residual are listed in Table 3 and
we find a mean slope of $\gamma = 1.16 \pm 0.14$.

Using a sharper turnover $\alpha = 1.5$ makes the fits slightly worse 
(the average of $\Delta_G$ is about 20 percent larger) and the 
best fit inner slopes are somewhat steeper $\gamma = 1.31 \pm 0.11$.
We also made some attempts with fitting procedures where $\alpha$ or $\beta$ 
or both $\alpha$ and $\beta$ are also free parameters. Like \citet{Klypin2001}
we found strong degeneracies, i.e. very different combinations of parameter
values can fit a typical density profile equally well. Therefore we only present
results from the fits with fixed $\alpha$ and $\beta$ parameters in this paper.

The fitting functions (\ref{Npro}) and (\ref{Gpro}) fit the measured density 
profiles very well over the whole resolved range. Function (\ref{Npro}) is even
a relatively good approximation {\it below} the resolved scale:
For example if one is extremely optimistic about $r_{\rm resolved}$ in run 
$D6$ and uses $r_{\rm resolved} = 2.8$ kpc instead of 13.5 kpc one gets
$\alpha_{\rm N} = 0.0203$,
$c_{\rm N} = 7.1$ and $\Delta_{\rm N} = 0.127$, while the
generalised fit is now clearly worse:
$\gamma_{\rm G} = 0.99$, $c_{\rm G} = 3.6$ and $\Delta_{\rm G} = 0.216$.
Also note that the residuals near $r_{\rm resolved}$ are 
very small or positive for (\ref{Npro}), i.e. the measured density
is as large as the fitted value. But at $r_{\rm resolved}$
it is possible that the measured density is slightly too low
since in this region the numerical limitations start to play a role.
If extrapolation beyond the converged radius is necessary it 
is not clear which profile is a safer choice. We agree with
\citet{Navarro2003} that all simple fitting formula have their 
drawbacks, that direct comparison with simulations should be attempted
whenever possible and that much higher resolution simulations
are needed to establish (or exclude) that CDM halos have 
divergent inner density cusps (as predicted in \citealt{Binney2003}).

\subsection{Maximum inner slope}

The results from the last section suggest that profiles
with a central cusp in the range $\gamma = 1.16 \pm 0.14$ 
provide a good approximation to the inner density profiles
of $\Lambda$CDM halos. But Figure 4 in \citet{Navarro2003}
seems to exclude our mean value for more than half of their 
cluster profiles. This is not totally inconsistent, but a hint
for a mild discrepancy that we will try to explain:
In principle the mass inside the converged radius limits the inner slope: 
$\gamma_{\rm max} = 3(1 - \rho(r)/\overline{\rho(,r)})$.
This is true if both the density and the cumulative 
density are correct down to the resolved scale.
But up to now the central density of a simulated profile 
always increased with better numerical resolution, so 
it is likely that also todays highest resolution simulations
underestimate the dark matter density near the centre. This
means that cumulative quantities 
like $v_{\rm circ}(r), \overline{M(<r)}$ and
$\overline{\rho(<r)}$ tend to be too low 
even at radii where the density has converged.
The converged radii used in \citet{Navarro2003} are 
close to the radius where the circular velocity is within 
10 percent of a higher resolution run, while the
density converges further in at about $0.6 r_{\rm conv}$ \citep{Hayashi2003}.
If we assume that this is also true for their highest resolution 
runs then $\overline{\rho(<r)} \propto v_{\rm circ}(r)^2$ is up to
20 percent too low, while the error in $\rho(r)$ is much smaller.
This raises the values for $\gamma_{\rm max}$ by about $0.2 \sim 0.3$
and our mean value $\gamma = 1.16$ is not excluded by any of 
their clusters anymore.
If the convergence with mass resolution is not as fast as 
$r_{\rm conv} \propto N^{-0.45}$ but rather $r_{\rm conv} \propto N^{-1/3}$,
see Section \ref{convergence tests}, then the maximum inner slopes could have
even larger errors.

\section{Comparison with other groups}

Recently, several groups have published simulations of dark matter 
clusters in the concordance cosmological model. 
These authors kindly supplied their density profiles and we 
show the comparison here. \citet{Fukushige2003}('F03') simulated four 
$\Lambda$CDM clusters with 7 to 26 million particles 
using a Treecode and the GRAPE hardware. These authors also used
the GRAFIC2 software \citep{Bertschinger2001} to generate their
initial conditions. 
\citet{Hayashi2003}('H03') and \citet{Navarro2003} presented eight clusters
resolved with up to 1.6 million particles within $r_{200}$ simulated with the
GADGET code \citep{Springel2001}, the method used to generate the
initial conditions is described in \citet{Power2003}. 
\citet{Tasitsiomi2003}('T03') simulated six clusters with up to
0.8 million particles within $r_{180}$ using the adaptive 
refinement tree code ART \citep{Kravtsov1997} and a technique
for setting up multi-mass initial conditions described in \citet{Klypin2001}.
\citet{Wambsganss2003}('W03') present a cosmological simulation without 
resimulation of refined regions, i.e. constant mass resolution 
(1024$^3$ particles in a 320 $h^{-1}$Mpc box). The four most
massive clusters in this cube are resolved within 0.5 to 0.9 million particles.
This simulation was performed with a Tree-Particle-Mesh (TPM) code 
\citep*{Bode2003} with a softening of 3.2 $h^{-1}$kpc.

\begin{figure*} 
\vskip 5.5 truein
\includegraphics{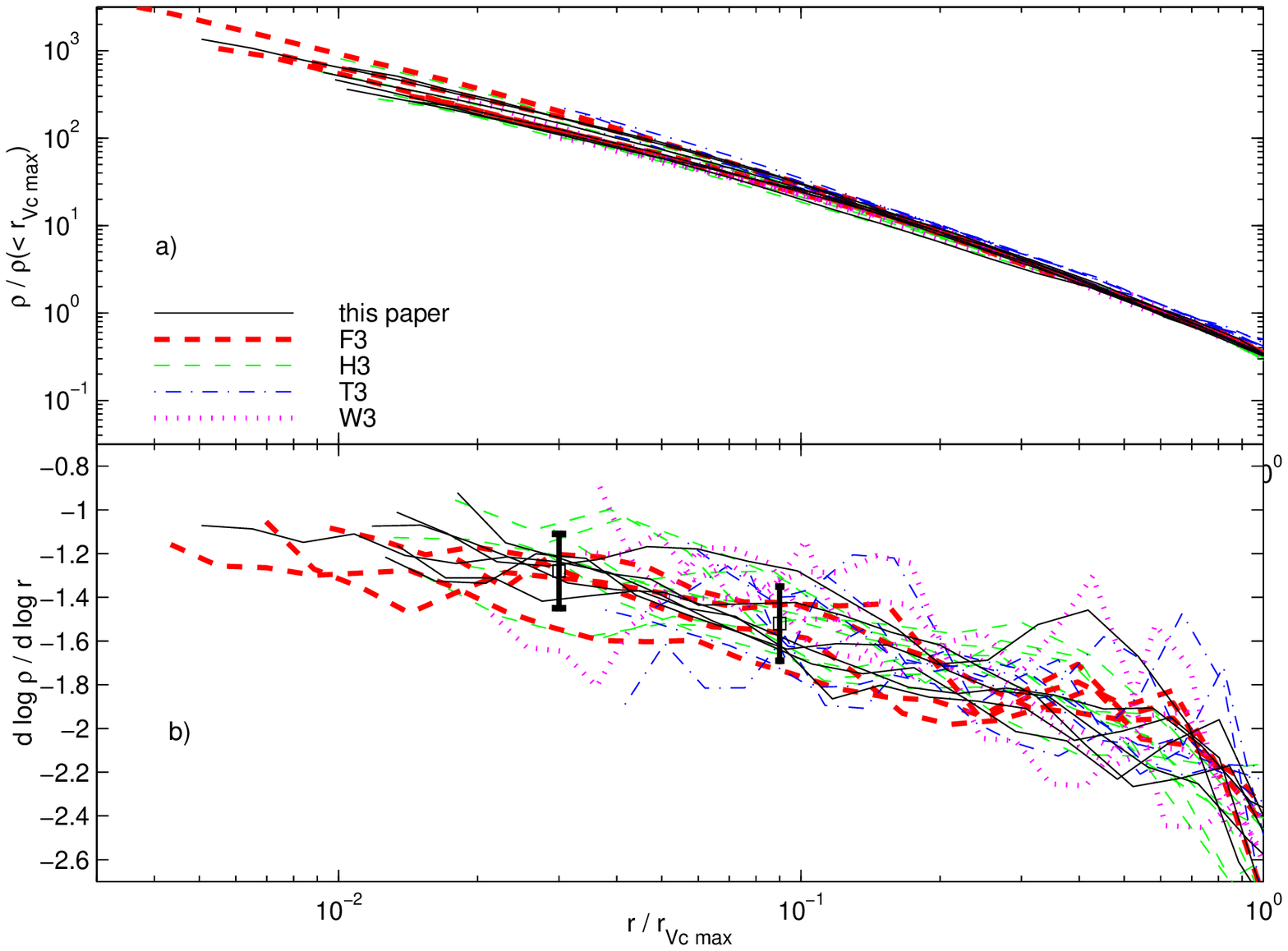}
\caption{\label{allPros.eps}Panel (a):
Density profiles of cluster simulated by different groups
Normalized to the radius were the circular velocity peaks $r_{\rm Vc max}$ 
and to $\overline{\rho(<r_{\rm Vc max})}$:
Six cluster from this paper (solid lines), 
four from \protect\citet{Fukushige2003} (thick dashed lines),
eight from \protect\citet{Hayashi2003} (thin dashed lines),
six from \protect\citet{Tasitsiomi2003} (dashed dotted lines)'
four from \protect\citet{Wambsganss2003} (dotted lines).
Despite the different
codes, parameters and initial conditions used the results are very similar.
Panel (b): Logarithmic slope for the profiles from (a). Points with
error bars give the averages at 0.03 and 0.09 
$r_{\rm Vc max}$ and a scatter of 0.15
(see Table \ref{tabSlopes}).} 
\end{figure*}

In Figure \ref{allPros.eps} we show these data along 
with the new simulations presented in this paper. We plot
the density profiles and the logarithmic slopes of the clusters all normalized 
at the radius 
such that the circular velocity curve peaks $r_{\rm Vc max}$
and to $\overline{\rho(<r_{\rm Vc max})}$. This corresponds to the radius at 
which $d log \overline{\rho}/ d log r = -2$.
We plot the curves to the ``believable'' radius stated by each group and 
down to about 0.01 $r_{\rm vir}$ for W03.

The density profiles are reassuringly similar. Furthermore, the scatter 
is small, the standard deviation of all profiles is 
roughly $\pm 0.15$ in the logarithmic gradient 
at small radii ($0.01 - 0.5 r_{\rm Vc max}$). 
Table \ref{tabSlopes} lists the measured slopes at different radii. There 
is no value at $3\% r_{\rm Vc max}$ for the cluster 
from \citet{Tasitsiomi2003} and \citet{Wambsganss2003} 
because this is below their quoted resolution limit.
Most values agree within the scatter, the profiles from \citet{Tasitsiomi2003}
are steeper when compared at 0.01 and 0.03 $r_{\rm vir} \equiv r_{98.4}$, 
but within the scatter at $3\% r_{\rm Vc max}$. This could be due to 
different halo selection. The majority of their clusters are not
isolated but in close pairs or triplets.
In a close pair the density falls slower with radius 
to 98.4 $\rho_{crit}$, so $r_{\rm vir} \equiv r_{98.4}$ 
is further out as in a isolated cluster
with similar inner profile. 
Among the samples of isolated clusters (F03; H03; W03 and our clusters) 
there is a small trend at 0.01 $r_{\rm vir}$ towards steeper slopes 
with better mass resolution. This could indicate that some numerical flattening
of the profiles is still present at 0.01 $r_{\rm vir}$ in the lower 
resolution clusters.

\begin{table}
\centering
\caption{Logarithmic slopes (absolute values) of our six high resolution
cluster density profiles. Line (a) gives the averages and scatter. (b)-(c) are
average slopes from other groups (see text for details).}
\label{tabSlopes}
\begin{tabular}{l | c | c | c | c }
  \hline 
  & $1\% r_{\rm vir}$ & $3\% r_{\rm vir}$ 
& $3\% r_{\rm Vc max}$ & $9\% r_{\rm Vc max}$ \\
  \hline
  $A9$&1.22&1.36&1.24&1.64\\
  $B9$&1.33&1.43&1.21&1.63\\
  $C9$&1.24&1.21&1.25&1.26\\
  $D12$&1.28&1.54&1.32&1.58\\
  $E9$&1.31&1.44&1.41&1.62\\
  $F9cm$&1.19&1.47&1.22&1.43\\
  \hline
  a) A-F&1.26$\pm 0.05$&1.41$\pm 0.11$&1.28$\pm 0.08$&1.53$\pm 0.15$\\
  b) F03
&1.25$\pm 0.05$&1.52$\pm 0.06$&1.33$\pm 0.15$&1.54$\pm 0.15$\\
  c) H03
&1.18$\pm 0.13$&1.38$\pm 0.14$&1.23$\pm 0.17$&1.50$\pm 
0.14$\\
  d) T03
&1.50$\pm 0.14$&1.79$\pm 0.07$&$ - $&1.56$\pm 0.12$\\
  e) W03
&1.11$\pm 0.04$&1.41$\pm 0.13$&$ - $&1.35$\pm 0.06$\\
  \hline
  avg.(a-e) &1.26&1.50& $-$ &1.49  \\
  avg.(a-c) &1.23&1.44&1.28&1.52  \\
  \hline
\end{tabular}
\end{table} 

\section{Summary}\label{Summary}

We have carried out a series of six very high resolution calculations of the 
structure of cluster mass objects in a hierarchical universe. The clusters 
contain up to 25 million particles and have force softening as small as 
$0.1\% r_{\rm vir}$.  

A convergence analysis demonstrates that for our Treecode with 
our integration scheme,
the radius beyond which we can trust the density profiles
scale according to the mean interparticle separation.
In the best case we reach a resolution of about $0.3 \%  r_{\rm vir}$.

Neither of the two parameter functions, the NFW and M99 profiles, 
are very good fits over the whole resolved range in most clusters.
One additional free parameter is needed to fit all six clusters:
The asymptotically flat profile from \cite{Navarro2003} and an NFW profile
with variable inner slope provide much improved fits. The 
best fit inner slopes are $\gamma=1.16\pm 0.14$.
{\it Below} the resolved radius the two fitting formulas used
are very different. Future simulations with
much higher resolution will show which one (if either) of the two is
still a good approximation on scales of $0.1\% r_{\rm vir}$
and smaller.

We compare our results with simulations from other groups who used
independent codes and initial conditions.  
We find a good agreement between the cluster density profiles
calculated with different algorithms. 
From $0.03 - 0.5 r_{\rm Vc max}$ the scatter in the profiles is nearly constant 
and equal to about 0.17 in logarithmic slope. At one percent of the virial radius
(defined such that the mean density within 
$r_{\rm vir}$ is $178 \Omega_M^{0.45} \rho_{\rm crit} = 
98.4 \rho_{\rm crit}$) the slope of the density profiles is $1.26 \pm 0.16$.

\section*{Acknowledgments}

We would like to thank the referee for many useful comments and
detailed suggestions.
We are grateful to Toshiyuki Fukushige, Eric Hayashi, 
Argyro Tasitsiomi and Paul Bode
for providing the density profile data from their cluster simulations.
We thank the Swiss Center for Scientific Computing in Manno for computing time
to generate the initial conditions for our simulations.
All other computations were performed on the zBox
supercomputer at the University of Zurich.
J. D. is supported by the Swiss National Science Foundation.

\bsp
\label{lastpage}
\end{document}